\documentclass{article}  
\usepackage{breckenridge}
\usepackage{graphicx}
\usepackage{wrapfig}
\usepackage{amssymb}

\newcommand{\hphm}{\mbox{$(\mathrm{h}^+ + \mathrm{h}^-)/2$}}
\newcommand{\npart}{\mbox{$N_{\mathrm{part}}$}}
\newcommand{\nbin}{\mbox{$N_{\mathrm{bin}}$}}
\newcommand{\dnchdeta}{\mbox{$dN_{\mathrm{ch}}/d\eta$}}
\newcommand{\nch}{\mbox{$N_{\mathrm{ch}}$}}

\newcommand{\sqrtsNN}{\mbox{$\sqrt{s_{_{\mathrm{NN}}}}$}}

\newcommand{\meanpt}{\mbox{$\langle p_T \rangle$}}
\newcommand{\pt}{\mbox{$p_T$}}

\newcommand{\gevc}{\mbox{${\mathrm{GeV/}}c$}}

\frompage{000} \topage{000}
\def\rightmark{Characteristics of \nch\ and \meanpt}
\def\leftmark{T.S.Ullrich}
 
\title{Characteristics of charged particle production in
    relativistic heavy-ion collisions}

\authors{
{Thomas S. Ullrich for the STAR Collaboration%
}\\[2mm]
{\normalsize
Brookhaven National Laboratory,
        Upton New York 11973-5000, USA\\[0.2ex] 
}}
 
\abstract {Inclusive spectra of charged particles at midrapidity in
    Au+Au collisions at $\sqrt{s_{_{NN}}}$ = 130 GeV and 200 GeV were
    measured with the STAR detector at RHIC.  The measured mean
    transverse momentum \meanpt\ shows a characteristic dependence on
    charged particle multiplicity and beam energy in Au+Au collisions
    that is distinctly different from $pp$, $p\bar{p}$ and
    $e^{+}e^{-}$ collisions. A $32\%\pm3\%$(syst) increase in \meanpt\ 
    from pp to Au+Au collisions was observed at 200 GeV. While the
    charged multiplicity was found to increase by $19\%\pm5\%$(syst)
    from $\sqrt{s_{_{NN}}}$ = 130 GeV to 200 GeV, no significant
    difference in \meanpt\ was found between the two energies. A
    comparison with model predictions is discussed.}

 
\begin{document}
 
\maketitle
\setcounter{page}{1}

\section{Introduction}

The systematic study of charged particle multiplicity in Au+Au
collisions at center of mass energies \sqrtsNN = 130 GeV and 200 GeV
was conducted by all RHIC experiments in great detail
\cite{qm2002,PHOBOS1,PHOBOS2,PHOBOS3,PHENIX1,BRAHMS1}.  These studies
were accompanied by intensive theoretical work successfully
connecting the produced particle densities with the initial conditions
in these collisions.  Essentially all models that attempt to describe
the evolution of matter produced in relativistic heavy-ion collisions
predict a strong correlation between the average transverse momentum
and the multiplicity of the event, typically $\meanpt^2 \propto
\dnchdeta$.  This simply reflects the basic thermodynamic concept
inherent to the models: large \nch\ means large entropy, large
entropy means higher temperature $T$, which, following the general
concepts of Landau's hydrodynamical model, implies an increase in
\meanpt. The use of thermodynamics in this argument 
does not, of course, preclude a dynamical description in terms of
(non-)perturbative Quantum Chromodynamics (QCD).  It simply takes
advantage of the fact that if the actual physical states of the system
are close to a canonical distribution, thermodynamics is an ideal tool
to analyze the measurements and relate them to bulk properties of hot
matter.

One prominent class of models that allows a direct coupling of global
observables with the initial conditions of the collision are those
based on the concept of gluon saturation. In these models particle
production is related to the initial gluon distribution. A semi-classic
calculation can be carried out due to the high gluon density, a
saturation phenomenon also known as the Color Glass Condensate
\cite{larry,DimaLevin}. The only scale that controls the gluon
distribution at a given Bjorken {\it x} is the saturation momentum
scale $Q_{s}$.  If the gluons realize themselves directly to hadrons,
the transverse momentum of the hadrons will be directly related to the
saturation scale $Q_{s}$ as well as the multiplicity of the particle
production\cite{DimaLevin,schaffner1}. In fact, this assumption was
used in some predictions that successfully predicted the particle
multiplicity along with their centrality and pseudorapidity dependence
\cite{DimaLevin}.

So far, however, only few studies were undertaken to verify if these
models are also able to describe the characteristics of \meanpt\ at
RHIC. In this article we present results on the characteristics of
\textit{(i)} the charged particle densities \textbf{and} \textit{(ii)}
of \meanpt\ using data recorded by the STAR experiment.


\section{Inclusive charged hadron spectra}

The data reported here were taken with the Solenoidal Tracker at RHIC
(STAR) \cite{stardet}. The main tracking device in STAR is a large
Time Projection Chamber (TPC) which provides momentum information and
particle identification for charged particles by measuring their
ionization energy loss.  A Central Trigger Barrel (CTB) constructed of
scintillator paddles surrounding the TPC and two Zero Degree
Calorimeters (ZDC) were used for triggering. Data were taken for Au+Au
collisions at \sqrtsNN=130 GeV in the summer of 2000 (run I) and
\sqrtsNN=200 GeV in the fall of 2001 (run II). The field setting for
run I was at 0.25 T while at run II it was at the nominal field
setting of 0.5 T. The increased field resulted in a factor of three
improvement in momentum resolution.  In both runs a minimum bias
trigger was defined by using coincidences between the two ZDCs located
$\pm 18$ meters away from the interaction point. For details see Ref.~\cite{STAR1}.

To measure the differential invariant cross-section of the produced
charged particle for different centralities tracks with transverse
momenta between $0.2$ and $2.0$ GeV/$c$ were selected.  The
reconstruction efficiency was determined by embedding simulated tracks
into real events at the raw data level, reconstructing the full
events, and comparing the simulated input to the reconstructed output.
Charged hadrons \hphm\ were approximated by the summed yields of
$\pi^{\pm}$, $K^\pm$, $p$ and $\bar{p}$. The efficiencies for the
identified particles are between 70\% and 80\% for central events at
$p_{T}>0.5$ GeV/$c$. At lower \pt\ (0.2 to 0.5 GeV/$c$), the
efficiency for pion is about 70\% and the efficiencies of kaon and
antiproton are from 30\% to about 70\% due to energy loss, absorption
and decay in the detector. The final efficiency of the charged
particle spectrum is the weighted efficiency of all $\pi^{\pm}$,
$K^{\pm}$ and $p^{\pm}$ according to their yield ratios measured at
$\sqrt{s_{_{NN}}}=130$ GeV and the proper Jacobian factor $\partial
y/\partial\eta(p_{T},\eta)$.  It is almost identical to that of the
pions for all \pt.

Background particles mis-identified as primary particles originate
from various sources. The dominant ones are: hadrons from weak decays
of $K_{S}^{0}$, $\Lambda$ and $\bar{\Lambda}$, weak decays of primary
charged particles, and hadrons from secondary interactions in the
detectors.  The background corrections for the former were calculated
by using measured yields and spectra.  The later two are important
only at low \pt\ and we used HIJING with a detailed detector
simulation to reproduce the background spectra for each centrality.

\begin{figure}[tb]
    \begin{center}
        \includegraphics[width=\textwidth]{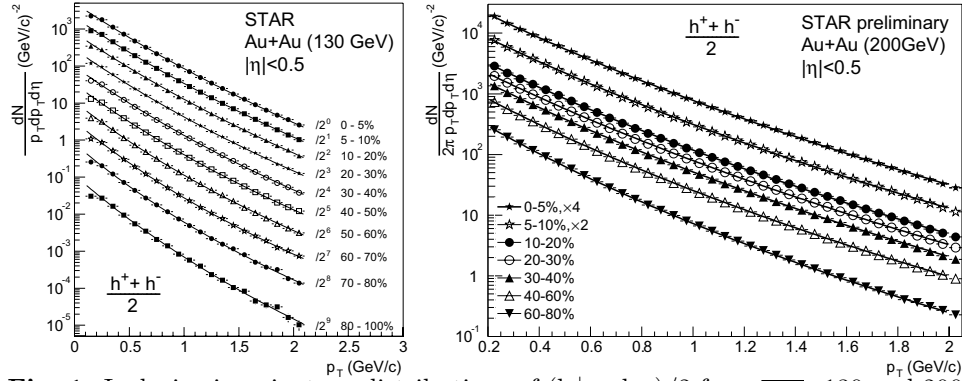}
\end{center}
\vspace{-9mm}
    \caption{Inclusive invariant \pt\ distributions of \hphm\ for
        \sqrtsNN=130 and 200 GeV Au+Au collisions.
        The lines are power law fits to the data (see text).}
    \label{fig:ptspectra}
    \vspace{-7mm}
\end{figure}
Figure \ref{fig:ptspectra} shows the fully corrected inclusive \pt\ 
distributions of primary charged particles for \sqrtsNN=130 and 200
GeV Au+Au collisions for 10 and 7 centrality bins, respectively.  The
spectra were fit with a power law function of the form $A
(1+\pt/p_0)^n$.  The average momentum is defined as $\meanpt =
2\;p_0/(n-3)$.  To extract the \meanpt\ we substitute $p_{0}$ by
$\meanpt (n-3)/2$ and fit the spectrum with \meanpt\ and $n$ as a free
parameters.  The difference between the integral of the fit function
and the sum of the data points in the fiducial kinematic range is less
than 1\%. The extrapolation to the total \pt\ is 21-26\% from central
to peripheral centrality. The systematical uncertainty of \meanpt\ 
was estimated to ~2\%.  The function, fit in the range of 0.2 \gevc\ 
to 2.0 \gevc, was then extrapolated to the full transverse momentum
range to get the total \nch. The systematical uncertainty on \nch\ is
~7\%.

For the 5\% most central collisions. The charged hadron multiplicity
at \sqrtsNN=200 GeV (130 GeV) is $d\nch/d\eta|_{|\eta|<0.5}=690 \pm
55(syst)$ ($579 \pm 60 (syst)$) and the mean transverse momentum
\meanpt\ is $0.517 \pm 0.012 \gevc$ ($0.519 \pm 0.01 \gevc$).  For
\sqrtsNN=200 GeV the multiplicity per participant pair increases by
38\% relative to $p\bar{p}$ and 52\% compared to nuclear collisions at
\sqrtsNN=17 GeV.  The \meanpt\ increases by 32\% and 20\%
respectively from $0.392$ GeV/$c$ in $p\bar{p}$ collisions and
$\simeq0.429$ GeV/$c$ in Pb+Pb collisions at SPS~\cite{STAR1,na49}.

\section{Centrality dependence of charged particle densities}

The RHIC experiments all have different methods of determining
centrality.  In general, all measure some final state variable (or
combination of these variables) that can be related to fractions of
the total measured hadronic cross-section.  These centrality bins are
then mapped to variables that allow for direct comparison between
experiments.  The primary variables that are used for comparison are
\textit{(i)} the number of participating nucleons in the collision
\npart\ and \textit{(ii)} the number of binary collisions \nbin, which
are determined via the application of the Glauber model phenomenology.
The basic concept of the Glauber model is to treat a Au+Au collision
as a superposition of many independent nucleon-nucleon (N+N)
collisions.  Thus, the only parameters that the model depends on are a
nuclear density profile (Woods-Saxon) and the non-diffractive
inelastic N+N cross-section.  The former is well measured in e+Au
scattering experiments.  The latter is well established by many
previous experiments.  With this, it is possible to determine both the
total number of independent N+N collisions (\nbin) and the total
number of nucleons that participate in the collisions (\npart) as a
function of impact parameter (b).  There are two separate
implementations of the Glauber model: Optical and Monte Carlo.  In the
optical formalism \nbin\ and \npart\ are directly determined by an
analytic integration of overlapping Woods-Saxon distributions.  In the
M.C. method an arbitrary number of events are simulated using a
computer program and the resulting distributions of \nbin\ and \npart\ 
are determined.  Once the distributions $d\sigma/d\npart$ and
$d\sigma/d\nbin$ are determined, these histograms are binned according
to fractions of the total cross-section.  This determines the mean
values of \nbin\ and \npart\ for each centrality class.  In this
context it should be noted that in the M.C. approach one requires that
all nucleons in either nucleus be separated by a distance $d \ge
d_{\mathrm{min}}$ where $d_{\mathrm{min}}=0.4$~fm is characteristic of
the length of the repulsive nucleon-nucleon force. Consequently the
total Au+Au cross section predicted by the Monte Carlo approach
depends strongly on $d_{\mathrm{min}}$. A value of
$d_{\mathrm{min}}=0.4$~fm yields a total cross-section of $7.2$~barns,
while $d_{\mathrm{min}}=0$~fm yields $6.8$~barns.
\begin{figure}[tb]
    \includegraphics[width=\textwidth]{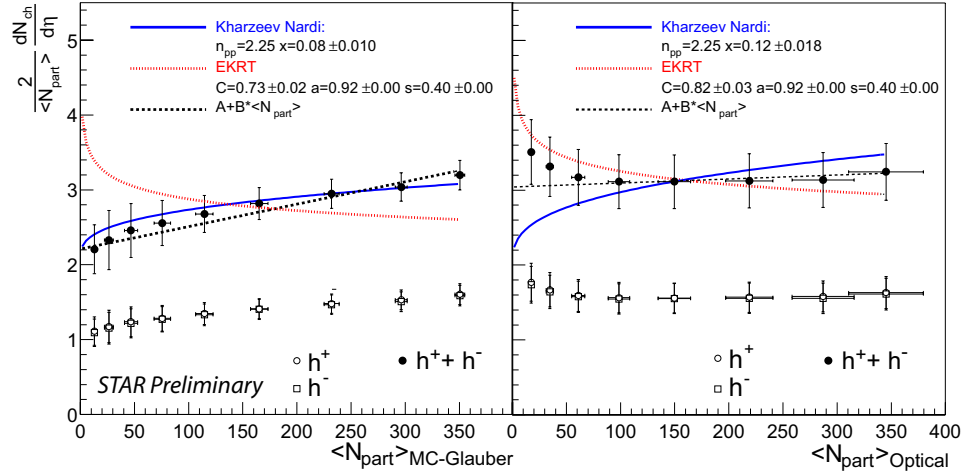} \vspace{-8mm}
    \caption{Rapidity density per participant nucleon pair
        $2/\langle\npart\rangle\ dN/d\eta$ versus number of
        participants $\langle\npart\rangle$.  On the left plot \npart\ 
        was calculated using a MC Glauber approach while on the right
        plot an optical Glauber model was used.}
    \label{fig:mcoptical}
\end{figure}

While at SPS energies both approaches yielded consistent answers, at
RHIC energies, the optical approach may
yield systematically smaller values of the total Au+Au cross-section.
Therefore, any results reported in terms of Glauber quantities must be
carefully interpreted based upon specifics of the Glauber
calculations.

In the case of STAR the centrality was determined by the multiplicity
of all charged particles within $|\eta|<0.5$ and then expressed in
terms of the fractional total hadronic cross section. The measured
data, i.e. the $dN/d\nch$ distributions, are then mapped to the
corresponding distribution obtained from Glauber calculations as
described above, thus relating \npart\ and \nbin\ to the measured
distributions.

Figure \ref{fig:mcoptical} shows the extracted yields as the
pseudo-rapidity density per participant pair $2/\npart\ dN/d\eta$
versus the number of participants \npart\ for Au+Au collisions at
\sqrtsNN=130 GeV, where for the left plot we used a Monte Carlo
Glauber calculation to derive \npart\ while for the right plot we used
an optical Glauber model.

The parameterization of pseudo-rapidity density of the EKRT model
\cite{EKRT}, a model based on the assumption of final state gluon
saturation, fails to describe the data when using the Monte Carlo
Glauber calculations, but agrees nicely with when using the
optical Glauber calculation. Both, our data and the EKRT
parameterization do not approach the charged particle pseudo-rapidity
density $n_{pp}=2.25$ (measured from non-single diffractive $p\bar{p}$
interactions and parameterized as $n_{pp} = 2.5\pm1.0
-(0.25\pm0.19)\ln(s) + (0.023\pm0.008)\ln^2(s)$ \cite{Abe1990}) in the
limit of \npart=2. One might argue that the model breaks down for very
peripheral collisions. The centrality dependence from mid-central to
central collisions is in fact too weak to rule out the EKRT model.

Comparing the two-component model described in Ref.~\cite{DimaNardi}
with our data shows the exact opposite behavior. Fitting the proposed
function to our data we find agreement between the fit and our
measurement only when using the Monte Carlo Glauber calculation (left
plot in Fig.~\ref{fig:mcoptical}). Here, we evaluate the fraction of
hard collisions $x_{\mathrm{hard}}$ as $0.08\pm0.01$ for our most
central events. The fraction of produced particles originating from
hard collisions is then calculated as $F_{\mathrm{hard}} = x\;n_{pp}\nbin/$
$(d\nch/d\eta) = 0.36 \pm 0.2$. Both  values $x_{\mathrm{hard}}$
and $F_{\mathrm{hard}}$ agree nicely with the
numbers quoted by Ref.~\cite{PHOBOS3} and \cite{DimaNardi}.

It should be noted that in general the M.C. Glauber method is
considered to be more accurate than the calculations in the optical
limit. Nevertheless, in both approaches the systematic uncertainties
in the calculation of \npart\ are considerable large for very
peripheral collisions, the region where the differences between the
various models become more apparent.  For the centrality fraction of
70-80\% the uncertainty on \npart\ already reaches 30\%.

\section{Characteristics of \meanpt\ in heavy-ion collisions}

Using the spectra from Fig.~\ref{fig:ptspectra} and the power law fits
described above we now turn to the centrality dependence of \meanpt.

\begin{wrapfigure}{r}{0.5\textwidth}
    \vspace{-5mm}
    \includegraphics[width=0.5\textwidth]{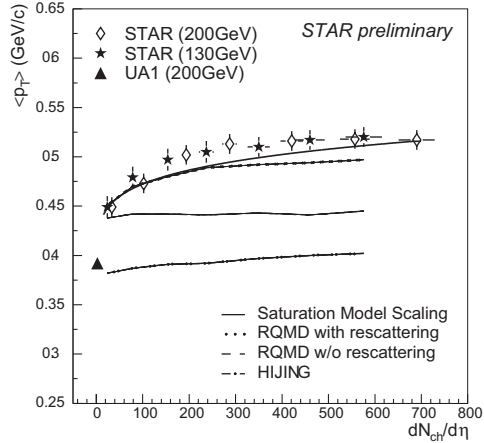}
    \vspace{0mm}
    \begin{minipage}[h]{0.53\textwidth}
     \caption{\meanpt\ as a function of \nch\ at \sqrtsNN = 130 GeV and 200 GeV.
         The solid line is a prediction assuming scaling of \meanpt\ 
         with \nch\ as described in the text. Also shown are the
         simulations from RQMD with and without rescattering and
         HIJING 1.35. All model curves are for \sqrtsNN = 130 GeV only.}
 \end{minipage}
\label{fig:ptscalingNch}
    \vspace{-2mm}
\end{wrapfigure}
Figure 3 shows the \meanpt\ as a function of
charge multiplicity for both energies. The open symbols are from 200
GeV data and the solid stars are measurements from 130 GeV data.  The
solid triangle depicts the UA1 measurement~\cite{ua1}.  We also show
the calculation from Eq.~\ref{eq:saturationpt1} discussed in more
details below. The simulations from RQMD with and without rescattering
are shown as dotted and dashed lines, respectively. The dot-dashed
line depicts the simulation from HIJING with default
settings. The \meanpt\ increases as function of centrality from
peripheral to semi-central collisions and seems to reach a plateau
for central collisions.

\begin{figure}[tbh]
\begin{minipage}[tb]{0.49\textwidth}
    \vspace{13mm}
    \includegraphics[width=\textwidth]{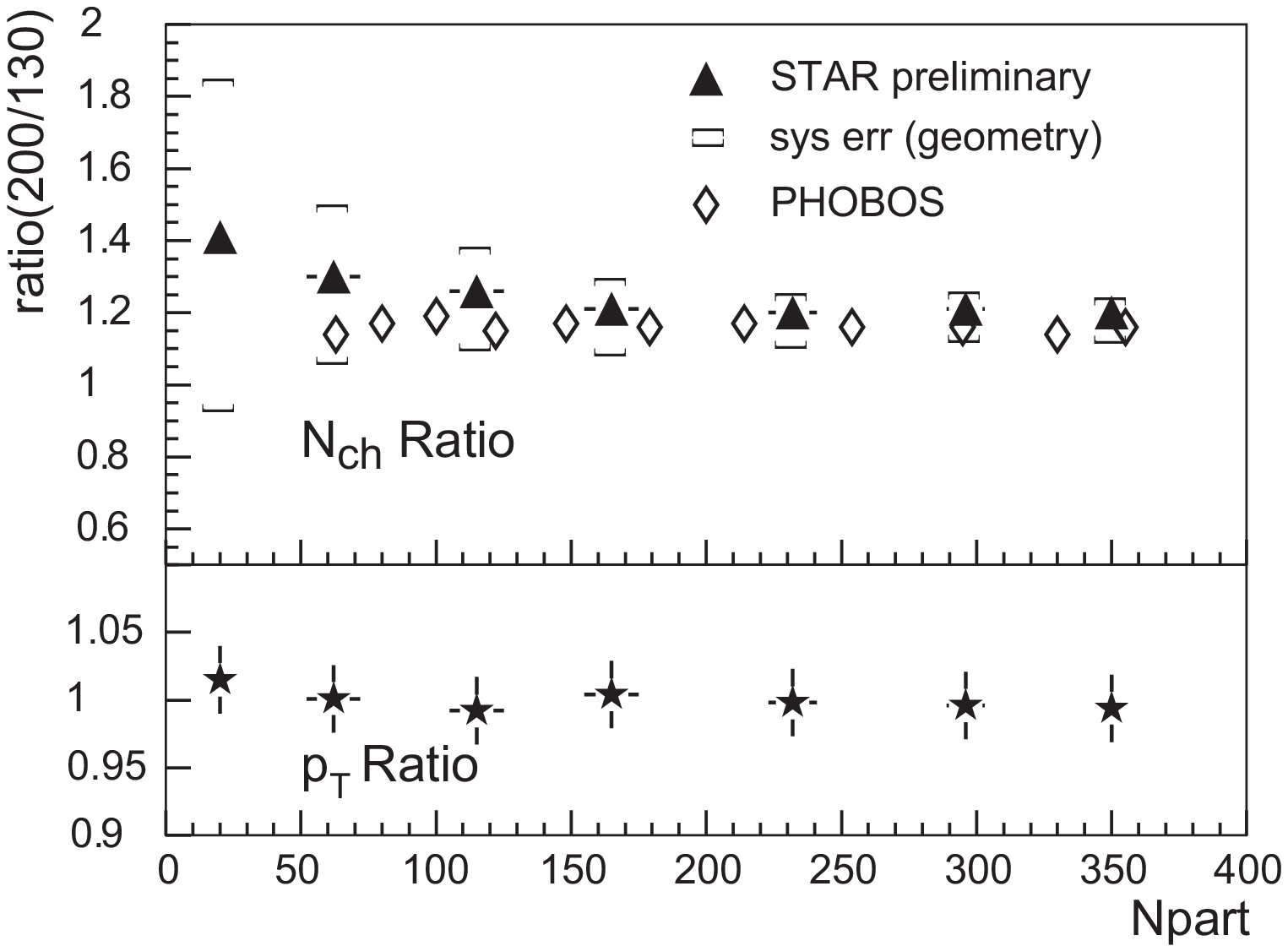}
    \vspace{-8mm}
    \caption{ Ratios of \nch\ and \meanpt\ between \sqrtsNN\ of 200
        GeV and 130 GeV Au+Au collisions. The open symbols are ratios
        measured by PHOBOS~\cite{PHOBOS2}.}
    \label{fig:ptNchRatios}
\end{minipage} 
\begin{minipage}[tb]{0.49\textwidth}
    \includegraphics[width=\textwidth]{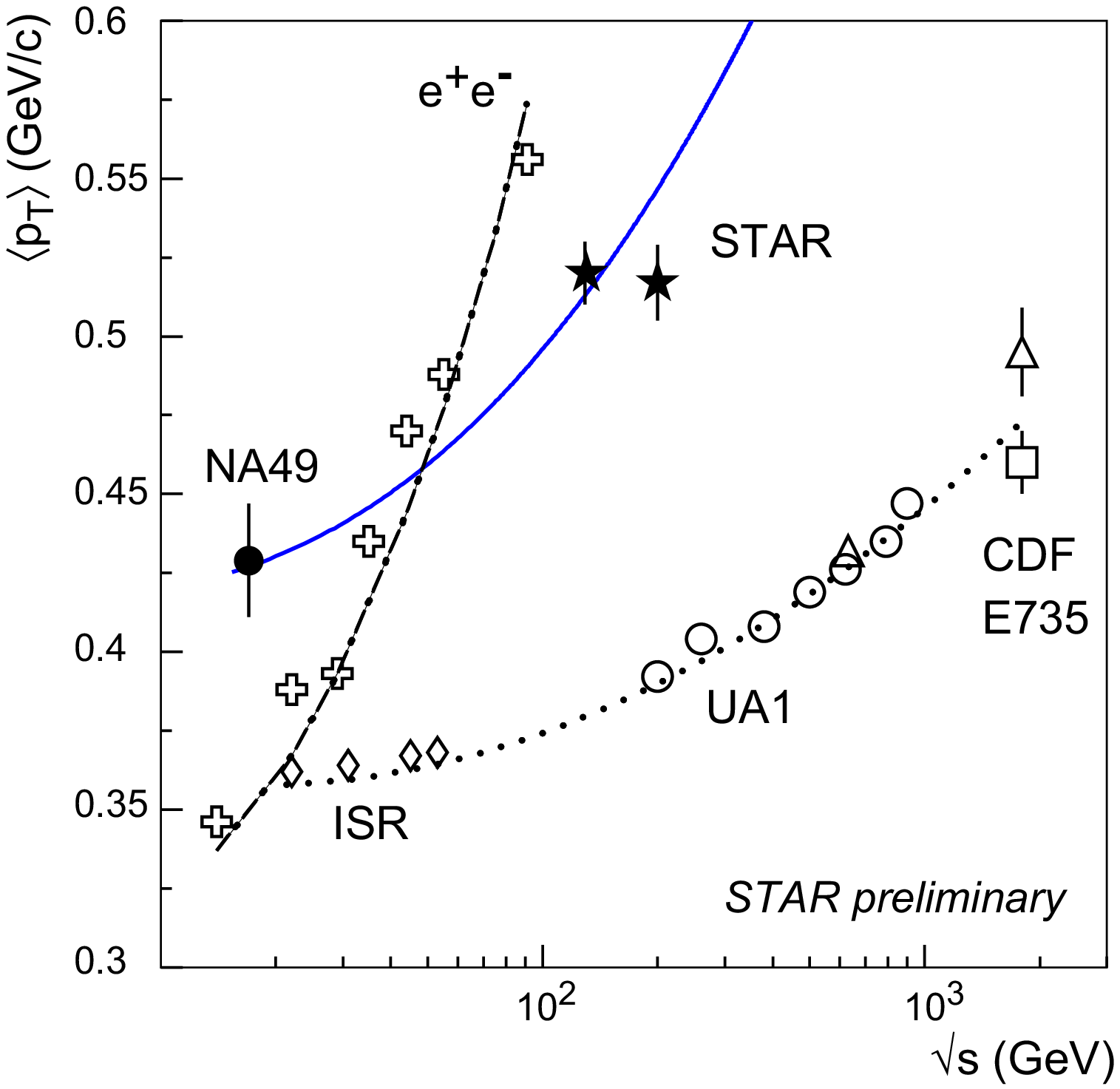}
    \vspace{-11mm}
    \caption{\meanpt\ of \hphm\  and h$^{-}$ as function of \sqrtsNN for
        $pp\ \bar{p}p$ and central AA collisions. The \meanpt\ from
        $e^{+}e^{-}$ was calculated along the jet thrust axis.}
    \label{fig:meanptS}
\end{minipage}
    \vspace{-5mm}
\end{figure}

Figure \ref{fig:ptNchRatios} has two panels. The solid triangles in
the top panel depict the inclusive charged particle ratios between 130
GeV and 200 GeV as a function of number of participants (neglecting
the small difference of \npart\ between these two beam energies).
The open diamonds are measurements from the
PHOBOS Collaboration~\cite{PHOBOS2}. For central collisions the
charged particle multiplicity increases from 130 GeV to 200 GeV
by 19\%. This result is
comparable with the results from other RHIC
experiments~\cite{PHOBOS2}. However, the systematical uncertainty at
peripheral collisions is too large to be conclusive as shown in the
top panel. Surprisingly, one finds no increase of \meanpt\ within the
current systematic errors for any centrality bin as depicted in the
bottom panel.

Figure \ref{fig:meanptS} shows \meanpt\ of negatively charged
particles $h^{-}$ from NA49 and charged particles \hphm\ from various
other experiments as function of \sqrtsNN\ for $pp,\ \bar{p}p$ and
central AA collisions.  The dotted line is a parametrization of the
measured \meanpt\ in pp collisions \cite{ua1}. For the $e^{+}e^{-}$
data, the dot-dashed line is a prediction from JETSET \cite{opal}.
The fact that the \meanpt\ from AA collisions is distinctly different
from both $pp$ and $e^{+}e^{-}$ indicates that AA collisions are not
simple superpositions of the elementary collisions.

As already discussed in the introduction, saturation~\cite{schaffner1}
and hydrodynamical~\cite{hydroscaling} models predict some scaling
behavior of $\meanpt^2 \sim {({{dN}\over{d\eta}})_{_{AA}}/{\pi{R^{2}}}}$.
\noindent

In order to get a quantitative analysis on
the dependence between \meanpt\ and multiplicity, we require that the
scaling dependence is linear and it should also satisfy the $p\bar{p}$
results.
Thereby, there is only one free parameter as shown in
Eq.~\ref{eq:saturationpt1}:
\begin{eqnarray}
\langle p_{T} \rangle_{_{AA}} = a+\sqrt{{s_{_{AA}}}\over{s_{pp}}}(\langle p_{T} \rangle_{pp}-a);
\label{eq:saturationpt1}
\end{eqnarray}
where $s_{_{AA}} = {({{dN}/{d\eta}})_{_{AA}}/{\pi{R^{2}}}}$ and
$s_{pp} = {({{dN}/{d\eta}})_{pp}/{\pi{r_{0}^{2}}}}$ are the
multiplicity densities per unit pseudorapidity per unit transverse area
in AA and pp collisions($R=r_{0}A^{1/3}$) and $a=0.3$ GeV/$c$ is a
constant chosen to describe the AA data.
For non-central collisions, the
calculation of the transverse area is not straight forward. We take
the centrality dependence parametrization of $s_{_{AA}}$ from
Ref.~\cite{schaffner1}. The solid curves in Fig. 3 and
Fig.~\ref{fig:meanptS} show the energy dependence and centrality
dependence of \meanpt\ from Eq.~\ref{eq:saturationpt1}. The curves
describe the data well from SPS to RHIC at 130 GeV. However, the data
show that the increase in \meanpt\ from \sqrtsNN = 130 to 200 GeV is
not as strong as the thick-line indicates which may invalidate the
scaling law from the saturation model \cite{schaffner1}.  On the other
hand, the energy dependence might indicate the importance of early
thermalization in the partonic stage, as proposed in ref
~\cite{mueller}.  Such initial partonic activity is consistent with
the early development of flow as indicated from $\langle v_2 \rangle$
measurements at RHIC \cite{starflow}. It will be interesting to see if
the description holds with an energy scan of lower beam energies at
RHIC.

\section{Summary}
In summary, STAR has measured charged particle multiplicities and
transverse momentum spectra at mid rapidity in Au+Au collisions at
\sqrtsNN=130 GeV and 200 GeV. Their centrality and beam energy
dependences are distinctly different from elementary collisions
($e^{+}e^{-}$, $pp$ and $\bar{p}p$).  The saturation model with direct
hadron production from gluons seems unable to explain the lack of
increase in \meanpt\ from 130 GeV to 200 GeV. However, the saturation
model with thermalization ~\cite{mueller} might be consistent with our
result. Further measurements with various beam energies and beam
species such as d+Au may be able to further distinguish different
production mechanisms.

\section*{Acknowledgment}
The author thanks F.~Laue, M.~Miller, and Z.~Xu for valuable contributions
for this article.

\vfill\eject

\begin{thebibliography}{99}  
  
\bibitem{qm2002} T.S.~Ullrich, Nucl.~Phys.~A715, 399c-411c (2003).
\bibitem{PHOBOS1} B.B.~Back \textit{et al.}, Phys.~Rev.~C65, 31901R (2002).
\bibitem{PHOBOS2} B.B.~Back \textit{et al.}, Phys.~Rev.~Lett.~88, 22302 (2002). 
\bibitem{PHOBOS3} B.B.~Back \textit{et al.}, Phys.~Rev.~C65, 061901R (2002). 
\bibitem{PHENIX1} K.~Adcox \textit{et al.}, Phys.~Rev.~Lett.~86, 3500 (2001).
\bibitem{BRAHMS1} I.G.~Bearden \textit{et al.}, Phys.~Lett.~B523, 227 (2001).
\bibitem{larry} L.~McLerran, Lect.~Notes~Phys.~583, 291 (2002); hep-ph/0104285.
\bibitem{DimaLevin} D.~Kharzeev and E.~Levin, Phys. Lett.~B523, 79-87 (2001).
\bibitem{schaffner1} L.~McLerran \textit{et al.},Phys.~Lett.~B514, 29 (2001);
                     J. Schaffner-Bielich \textit{et al.}, nucl-th/0202054.
\bibitem{stardet} C.~Adler \textit{et al.}, Nucl.~Instr.~Meth.~A499, no.~2+3 (2003).
\bibitem{STAR1} C.~Adler \textit{et al.}, Phys. Rev. Lett.~87, 112303 (2001).
\bibitem{na49} H.~Appelsh\"auser \textit{et al.}, Phys.~Rev.~Lett.~82, 2471 (1999).
\bibitem{EKRT} K.~Eskola \textit{et al.}, Nucl.~Phys.~B570, 379 (2000); hep-ph/9909456.  
\bibitem{Abe1990} F.~Abe \textit{et al.}, Phys.~Rev.~D41, 2330 (1990).      
\bibitem{DimaNardi} D.~Kharzeev and M.~Nardi, Phys.~Lett.~B507, 121-128 (2001).
\bibitem{ua1} C.~Albajar \textit{et al.}, Nucl.~Phys.~B335, 261 (1990).
\bibitem{opal} R.~Akers \textit{et al.}, Phys.~Lett.~B320, 417(1994).
\bibitem{hydroscaling} D.~K.~Srivastava, Phys.~Rev.~C64, 06490 (2001).
\bibitem{mueller} A.~Mueller \textit{et al.}, Nucl.~Phys.~A715, 20c (2003);
                  R.~Baier \textit{et al.}, Phys.~Lett.~B539, 46 (2002).
\bibitem{starflow} K.H.~Ackermann \textit{et al.}, Phys.~Rev.~Lett.~86, 402(2001);
                   C.~Adler \textit{et al.}, Phys.~Rev.~Lett.~87, 112301 (2001);
\end{thebibliography}
\end{document}